# Annihilation mechanism of excitons in a MoS$_2$ monolayer through direct Förster-type energy transfer and multistep diffusion

Kwang Jin Lee[1], Wei Xin[2], and Chunlei Guo[1,2]*

[1] Institute of Optics, University of Rochester, Rochester, New York, USA,

[2] GPL, Changchun Institute of Optics, Fine Mechanics and Physics, Chinese Academy of Sciences, Changchun, China, 130033

*Correspondence author E-mail: guo@optics.rochester.edu

**Atomically thin MoS$_2$ layer is a direct bandgap semiconductor exhibiting strong electron-hole interaction due to the extreme quantum confinement and reduced screening of Coulomb interactions, which results in the formation of stable excitons at room temperature. Therefore, various excitonic properties of MoS$_2$ monolayer are extremely important in determining the strength of light–matter interactions including their radiative recombination lifetime and optoelectronic response. In this paper, we report a comprehensive study of the underlying annihilation mechanism of various types of exciton in MoS$_2$ monolayer using the transient absorption spectroscopy. We rigorously demonstrate that the Förster-type resonance energy transfer is the main annihilation mechanism of A- and B-excitons while multistep diffusion process is responsible for C-exciton annihilation, which is supported by critical scientific evidence.**

## I. INTRODUCTION

Two-dimensional (2D) transition metal dichalcogenides (TMDs) have emerged as fascinating materials for low-dimensional applications owing to their remarkable electronic and optical properties [1–5]. These materials are highly attractive for fundamental studies of novel physical phenomena and for applications ranging from nanoelectronics and nanophotonics to sensing and advanced optoelectronic devices [5–8]. For example, their unique electronic band structure exhibits a lot of remarkable characteristics, such as gate-tunable conductivity with relatively high mobility [9], strong photoresponse [10], and valley-selective optical excitation [11–15]. Based on these properties, novel functional devices have been developed, including field-effect transistors (FETs) [9,16–18], broadband photodetectors [19], light-harvesting devices [20], photosensor [21], chemical sensor [22], and valleytronics [23–25].

2D TMDs are 2D semiconductors with direct bandgap lying in the visible and near-IR range at the energetically degenerate K and K´ (−K) points of hexagonal Brillouin zones, enabling strong interactions of dipole transitions with light [26]. Quantum confinement effect and reduced dielectric screening lead to strong Coulomb interactions between electrons and holes, resulting in tightly bound excitons with large binding energy (0.2 ~ 0.8 eV) [27–29]. Therefore, novel applications based on 2D TMDs can be achieved by systematically understanding the excitonic properties such as excitonic band structure, migration dynamics, and multiexcitonic states.

Monolayer molybdenum disulfide ($MoS_2$), one of the members of 2D TMD materials, is a direct bandgap semiconductor with strong photoluminescence [26,30]. The excitonic property in monolayer $MoS_2$ is mainly dependent on the valence-band splitting due to strong spin-orbit coupling, which leads to Coulomb-enhanced multiexciton excitations at the bandedge (K and K' points), so-called the A- and B- excitons. Hence, various excitonic properties of A- and B-excitons such as exciton-absorption bleaching, interexcitonic interaction, and broadening has been intensively studied [31]. In addition, due to the parallel bands in their density of states, $MoS_2$ monolayer shows strong optical responses for excitation energies higher than bandgap [32,33]. This photoexcited exciton in the 'band nesting' region, denoted as C-exciton, exhibits a fast intraband relaxation and a very slow indirect emission process arising from spontaneous charge-separation in the momentum space [34,35]. Hence, the C-exciton is delocalized to overlap with the continuum states near the K (−K) point in the Brillouin zone, leading to completely different hot-carrier relaxation process compared with the band-edge excitons (A- and B- excitons) [35,36]. It can also result in a strong optical response in absorption spectrum as with band-edge excitons.

The excitonic properties in 2D TMDs are extremely important in determining the confinement-enhanced characteristics. For example, exciton migration dynamics in $MoS_2$ monolayer is of high relevance for applications such as light harvesting systems and light emitting devices [37,38]. At high excitation intensities, a many-body interaction where one exciton is annihilated by transferring its energy to another exciton, so-called exciton-exciton annihilation (EEA), can take place in 2D TMDs and organic semiconductors [39–48]. EEA usually occurs when two excitons are sufficiently close to interact and to generate a single exciton with a higher energy. This indicates that EEA is the additional deactivation process of excitons via interaction of two of them and is known to strongly affect the performance of light-emitting diodes at high excitation densities. A critical step in the operation of solar cells and photodetectors is also highly associated with EEA. For instance, the exciton diffusion length, the distance over which an exciton can migrate before it decays within its lifetime, is one of the most important parameters to optimize the photocurrent, which is dictated by EEA process. In this context, comprehensive understanding of exciton annihilation and the accurate determination of the exciton diffusion length is essential for the optimization of photovoltaic device structures. Therefore, EEA is a crucial phenomenon to figure out underlying mechanism of exciton annihilation in 2D semiconductors, organic polymers, carbon nanotubes, and molecular optical switches.

In general, EEA takes place through two distinct mechanisms, a direct Förster-type resonance energy transfer (FRET) and multistep diffusion [44]. FRET occurs in a way that direct long-range energy transfer via dipole-dipole interaction gives rise to annihilation, which depends primarily on the overlap between the emission spectrum of the donor and the absorption spectrum of the acceptor. In case of two identical excitons, a spectral overlap means the energy overlap between the exciton emission and the excited state absorption (ESA, absorption from the exciton state to higher electronic states), resulting in annihilation step, $E_1 + E_1 \rightarrow E_n + E_0 \rightarrow E_1 + E_0$ where $E_0$ and $E_1$ are ground state and exciton state, respectively. $E_n$ is the final state upon photoexcitation. As the highly excited state $E_n$ generated by FRET relaxes quickly to exciton state $E_1$, and thereby quenching efficiently one exciton [43]. On the other hand, the diffusion process induces a continuous exciton annihilation through the multiple transfer steps between the exciton and the ground state. [43,49]. Consequently, the diffusion model assumes that the excitons move like random walkers in many steps towards each other of the type $E_1 + E_0 \rightarrow E_0 + E_1$, until they annihilate via a short range interaction in a final step of the type $E_1 + E_1 \rightarrow E_n + E_0 \rightarrow E_1 + E_0$.

Recently, control of exciton dynamics in $MoS_2$ monolayer has been investigated through optical interplay with hyperbolic metamaterials [50]. The study found that Förster radius and FRET efficiency were enhanced by nonlocal effect from hyperbolic metamaterials. However, a more comprehensive study is required to understand why FRET and diffusion processes are dominant annihilation mechanisms for A- and C-excitons, respectively. In this article, we fully identify and characterize the annihilation of A-, B-, and C-excitons in a $MoS_2$ monolayer by employing ultrafast transient absorption spectroscopy. We demonstrate that A- and B-excitons are predominantly governed by FRET process while the diffusion process entirely dictates the annihilation of C-exciton, which is corroborated by several critical evidences.

## II. SAMPLE AND EXPERIMENTAL METHODS

The preparation of a single-layer $MoS_2$ on silicon substrate (<100>, ≈300 nm $SiO_2$) is based on the traditional chemical vapor deposition (CVD) method (high temperature, Argon environment for 2.5 h). $MoO_3$ (99.99%, Aladdin) and S (99.99%, Alfa Aesar) powders were chosen as precursor materials. The PTAS (perylene-3, 4, 9, 10-tetracarboxylic acid tetra potassium salt) was also dropped on the substrate as the seeding promoter to increase the nucleation. To remove the surface contaminants and make a close contact between $MoS_2$ and the substrate, annealing at 300 °C for 1 h was used. Figure 1a shows the absorption and photoluminescence spectra of $MoS_2$ monolayer. The two absorption peaks at 1.87 eV and 2.05 eV correspond to A- and B-excitons, respectively. The broad absorption band above 2.80 eV corresponds to the C-excitons. The photoluminescence (PL) peak and shoulder at 1.84 eV and 2.01 eV are responsible for A- and B-excitons, respectively. No photoluminescence was observed for C-excitons. $MoS_2$ monolayer was characterized using Raman spectroscopy and displayed the expected Raman modes $A_{1g}$ and $E_{2g}$, with ~ 20 $cm^{-1}$ separation for the monolayer area as shown in Figure 1b. Figure 1c and 1d display the Raman signal intensity map of the sample. We can clearly see the Raman signal for $A_{1g}$ and $E_{2g}$ modes at 382 $cm^{-1}$ and 402 $cm^{-1}$, respectively, which is consistent with the previous work [51]. It is also worth noting that we observed neither noticeable PL feature in the lower energy (Figure 1a) nor significant change in Raman spectra (Figure 1b) due to defect states such as S vacancy and oxygen impurities.

To investigate the EEA, we employ ultrafast transient absorption (TA) spectroscopy based on the femtosecond pulse laser, which is incredibly useful tool to investigate the carrier dynamics, photosynthesis, and charge transfer dynamics in semiconducting materials [52–57]. A Ti:sapphire regenerative amplifier system operating at a repetition rate of 1 kHz and delivered 67-fs pulse duration centered at 800 nm was used. The laser output is split into a pump and a probe beam by a beam splitter. For the A- and B-excitons, the pump and probe beams were chosen by using a 5-mm thick sapphire window for generating white-light-continuum, followed by band-pass filters centered at 2.25 eV for the pump and 1.85 eV (A-exciton) and 2.01 eV (B-exciton) for the probe. Probe beam energy was chosen by the PL peak energy of A- and B-excitons, 1.85 eV and 2.01 eV, respectively. For the C-exciton, second harmonic of fundamental beam (using a BBO crystal) was chosen, which was further spilt into a pump and a probe. The pump beam was modulated using a mechanical chopper at 220 Hz and the relative reflection $\Delta R/R$ of the probe beam as a function of the time delay was further read out with a photodiode. The relative reflectance is given by $\Delta R/R = (R_{on} - R_{off})/R_{off}$, where $R_{on}$ and $R_{off}$ are the sample reflectance with the pump beam on and off, respectively. In this study, we used linear polarized beam in entire experiment, allowing us to rule out valley-dependent effect.

### III. THOERETICAL MODEL

In this section we discuss the EEA theoretically by considering exciton self-quenching dynamics. Typically, high excitation intensity or pump fluence is required to induce EEA, which may also induce other nonlinear phenomena such as Auger recombination. At higher excitation intensities, the exciton decay becomes more rapid at an initial time range but independent of the intensity at a longer time range. The faster decay strongly depends on the excitation intensity (or pump fluence), exhibiting higher order reactions resulting from EEA, while the slower decay corresponding to intrinsic exciton lifetime via the radiative and nonradiative deactivations process is independent of excitation intensity. TA kinetics for each exciton in the initial time range can be analyzed by the EEA formalism given by the rate equation as following [47−49]

$$\frac{d}{dt}n(t) = -\frac{n(t)}{\tau} - \frac{1}{2}\gamma(t)n(t)^2 \tag{1}$$

where $n(t)$ is the exciton density at a delay time $t$, $\gamma(t)$ is the bimolecular annihilation rate coefficient proportional to $t^{-1/2}$ and $\tau$ is intrinsic the exciton lifetime at the low exciton density limit. The factor 1/2 represents that only one exciton is left after EEA ($E_1 + E_1 \rightarrow E_n + E_0 \rightarrow E_1 + E_0$ + phonon). In general, photoexcitation in conventional inorganic semiconductors such as III−V quantum wells generates free carriers, rather than excitons at room temperature. Therefore, the corresponding annihilation process can normally be described as a three-body Auger recombination process, a nonradiative decay mechanism commonly shown in highly excited semiconductors with free charge carriers. Indeed, exciton annihilation is generally analogous to the Auger recombination process. However, owing to the high exciton binding energy, bi-excitonic interaction is dominant over the Auger process in monolayer $MoS_2$, and thus the Auger constant that is proportional to $n(t)^3$ can be neglected in this study [39]. Given that a typical III−V materials with band gap comparable to the A-exciton energy in monolayer $MoS_2$ possess extremely small Auger coefficient [58], many-body effects play a much more dramatic role in EEA processes in a $MoS_2$ monolayer than in conventional semiconductor systems [39]. Again, this difference is attributed to the strong confinement effects in 2D nature and the associated reduced dielectric screening.

In Eq. (1), $\gamma(t)$ is given by Eq. (2a) and (2b) for the FRET and diffusion models, respectively, as follow,

$$\gamma_F(t) = \frac{R_F^2}{2}\sqrt{\frac{\pi^3}{\tau t}} \tag{2a}$$

$$\gamma_D(t) = \frac{1}{aN_0}\sqrt{\frac{8D}{\pi t}} \tag{2b}$$

where $R_F$ and $D$ are the Förster radius and diffusion coefficient, respectively. $a$ and $N_0$ are the lattice constant and molecular density, respectively.

## IV.  DIFFUSION PROCESS: C-EXCITONS

As mentioned earlier, the parallel band region can promote self-separation of C-excitons to generate hot carriers in momentum space, and they rapidly relax to nearest band extrema, $\Lambda$ and $\Gamma$ valley [35]. Recombination of these hot carriers from C-excitons cannot generate photons owing to momentum mismatching, and they generally would release excess energy in the form of phonons. Because of this, even though the C-exciton may not play a critical role in light-harvesting applications, comprehensive understanding C-exciton dynamics is essential to collecting the high-energy hot carriers [36].

Figure 2a shows the TA kinetics of C-excitons in MoS$_2$ monolayer for different pump fluences ($f_{\text{pump}}$ = 1, 5, 15, 25, 50 μJcm$^{-2}$) at 3.05 eV with probe energy of 3.05 eV. A positive signal of $\Delta R$ was attributed to a reduction of the available ground state carriers due to excitation from the pump, so-called a ground state bleach. Since the C-excitons will not remain in parallel band for a long time due to ultrafast self-separation, the photobleaching signals (amplitude of $\Delta R/R$) of C-exciton present a weaker signal compared to band-edge excitons (Figure 3). However, weak TA signals of C-exciton cannot influence to acquire the decay dynamics of C-exciton. At the lowest $f_{\text{pump}}$, TA kinetics shows single decay time ($\tau$) of 219 ps, which corresponds to the intrinsic exciton lifetimes. As $f_{\text{pump}}$ increases, additional decay channel of C-exciton caused by EEA comes into play and leads to another shorter decay time constant formed in the initial delay time range (up to ~ 100 ps). We find that the short time constant ($\tau_1$) decreases with $f_{pump}$ while intrinsic exciton lifetime ($\tau_2$) is independent of $f_{pump}$ (Table 1). We note that $\tau_1$ is responsible for EEA. Figure 2b exhibits the linear relationship between initial magnitude of TA signal ($\Delta R/R$) and $f_{pump}$. Typically, at higher excitation levels, the bleach of the absorption band can also contribute to TA signal, resulting in a $\Delta R/R$ that is no longer linear with $f_{pump}$. In this work, excitation intensities are all controlled to be in the linear region, so the photo-induced exciton absorption bleach dominates the TA response [59]. For C-excitons, FRET can be completely ruled out owing to their non-emissive property [49]. In order to determine the diffusion coefficient $D$, TA kinetics can be fitted by the solution of Eq. (1) for 1D diffusion, which is given by [48−50]

$$n_{1D}(t) = \frac{n_0 e^{-t/\tau}}{1 + \frac{n_0}{aN_0}\sqrt{2D\tau}\,\mathrm{erf}\left(\sqrt{\frac{t}{\tau}}\right)} \quad (3)$$

where 'erf' is the error function. $n_0$ is the initial exciton density right after photoexcitation determined by the relation $n_0 = f_{pump}(1-10^{-A})/E_{pump}$, where $A$ and $E_{pump}$ are the absorption coefficient and energy of pump beam, respectively [49]. Here, the exciton lifetime $\tau$ was kept as a constant (213 ps). Black solid lines in Figure 2a are the fitting curves with Eq. (3). On the basis of values of $a$ and $N_0$ of $MoS_2$ monolayers, which are 3.16 Å and $5.7 \times 10^{14}$ cm$^{-2}$, respectively, we extracted the values of $D$ for each $f_{pump}$ and plot them as a function of inverse square root of the pump fluence, $(f_{pump})^{-1/2}$ in Figure 2c. Since the temperature ($T$) induced by pump is proportional to $f_{pump}$, linear relationship between $D$ and $(f_{pump})^{-1/2}$ indicates that $D$ linearly increases with inverse square root of temperature ($T^{-1/2}$) [60]. In the previous work, for free excitons where the exciton jumps before the lattice relaxes around the excited molecule, the relationship $D \sim T^{-1/2}$ was derived for the situation where the scattering by phonons is dominant mechanism limiting the mean free path of the exciton [60]. Therefore, the linear behavior of $D$ with $T^{-1/2}$ clearly indicates that the diffusion process of C-exciton takes place through exciton-phonon coupling. Figure 2d presents the annihilation rate with diffusion process, $\gamma_D(t)$, as a function of time delay (Eq.(2a)) for different pump fluences. The higher pump fluence, that is, the higher exciton density, the faster annihilation rate.

## V. FRET PROCESS: A- AND B-EXCITONS

Figures 3 presents the TA kinetics of A- and B-excitons in $MoS_2$ monolayer for the same pump fluences ($f_{pump}$ = 1, 5, 15, 25, 50 μJcm$^{-2}$) at 2.25 eV with probe energy of 1.85 eV and 2.01 eV, respectively. Positive signal of $\Delta R$ for A- and B-excitons may be attributed to either a ground state bleach or stimulated emission of the pump-induced excited states. Similar to C-exciton, a single decay time (intrinsic exciton lifetimes) of 189 ps (181 ps) was obtained for A- (B-) exciton at the lowest $f_{pump}$. Relatively short exciton lifetime in these band edge excitons compared to C-exciton is indeed consistent with previous works in which a longer lifetime of C-excitons was reported owing to favorable band alignment and transient excited state Coulomb environment [51]. In addition, as we mentioned earlier, no feature arising from the defect state was observed in photoluminescence and Raman spectra, which indicates that defect states should be nonradiative even if they can be excited at all. However, no other time constants responsible for possible defect states were observed at the lowest exciton density (black dots in Figure 3a and 3b), which plainly means that defects do not play a significant role in this study.

We also notice that EEA comes into play for both A- and B-excitons as $f_{pump}$ increases, giving rise to the emergence of short time constant ($\tau_1$) (Table 1). The longer time constant ($\tau_2$) that is independent of $f_{pump}$ corresponds to the intrinsic exciton lifetime. It is worth noting that TA kinetics of A- and B-excitons exhibit stronger dependence on $f_{pump}$ compared to C-exciton, which is consistent with the efficient dissociation of C-exciton due to self-separation of photocarriers in the band nesting region [35]. Since the TA kinetics of A- and B-excitons are almost identical, we will analyze the A-exciton results only in the following.

In general, diffusion coefficient $D$ for exciton annihilation may or may not be dependent on $f_{pump}$. In addition, diffusion process can take place regardless of emissivity of exciton. On the other hand, the FRET occurs only when exciton is an emissive and is independent of $f_{pump}$. In this section, we discuss the FRET as a possible annihilation mechanism of A- and B-excitons. Similar to diffusion, the solution of Eq. (1) for FRET in 2D materials is given by [50]

$$n_F(t) = \frac{n_0 e^{-t/\tau}}{1 + \frac{n_0 R_F^2 \pi^2}{4} \mathrm{erf}\left(\sqrt{\frac{t}{\tau}}\right)} \quad (4)$$

where the $R_F$ is the Förster radius, the distance between donor and acceptor at which the energy transfer efficiency is 50%. $R_F$ depends on the overlap integral of the donor emission spectrum with the acceptor absorption spectrum and their mutual orientation as expressed by the following equation [61].

$$R_F^6 = \frac{9\eta_D \kappa^2}{128\pi^5 n^4} \int d\lambda \lambda^4 F_D(\lambda) \sigma_A(\lambda) \quad (5)$$

where $\kappa$ is the dipole orientation factor, $n$ is the refraction index of the environment, $\lambda$ is the wavelength, $F_D$ is the normalized emission spectrum of donor, and $\sigma_A$ is the absorption cross-section of acceptor. $\eta_D$ is the quantum yield of isolated donor expressed as the ratio of the rate of radiative recombination to the total rate of exciton decay. We note that $\eta_D$, $F_D$ and $\sigma_A$ are the individual properties for isolated configuration so that the above equation is not associated with the density of donor and acceptor. Black solid lines in Figure 3a and 3b are fitting curves based on the Eq. (4), from which we extract the values of $R_F$.

First, let us assume that A-exciton annihilates via diffusion. According to previous study [60], the temperature dependence of the diffusion coefficient for free excitons can be either proportional to $T^{-1/2}$ or constant depending on whether the temperature is above the Debye temperature ($T_D$) or not. In other words, $D$ is constant for $T < T_D$ and has $T^{-1/2}$ dependence for $T > T_D$. Since we have already demonstrated that $D_C$ represents $T^{-1/2}$ dependency (Figure 2c), we can conclude that the temperature of C-exciton ($T_C$) is higher than $T_D$, ($T_C > T_D$). For A-exciton, we can also plot the $D_A$ versus $(f_{pump})^{-1/2}$ under this assumption. As shown in Figure 4, the constant behavior of $D_A$ (blue), indicates that diffusion of A-exciton takes place via exciton-phonon coupling under $T_A < T_D$. Therefore, this analysis unambiguously shows that $T_C$ is higher than $T_A$ ($T_A < T_C$). However, the same $f_{pump}$ for A- and C-excitons indicates $T_A = T_C$, which is obviously inconsistent with $T_A < T_C$. Consequently, our assumption that A-exciton annihilates via diffusion process is incorrect. Given the fact that there are only two mechanisms of exciton annihilation, FRET and diffusion, this systematic analysis, proving that the diffusion cannot be the annihilation mechanism of A-exciton, is very strong evidence for underlying annihilation mechanism of A-exciton being a FRET, not a diffusion.

Second, we extract $R_F$ values by fitting TA kinetics shown in Figure 3a with Eq. (4) and plot it versus $(f_{pump})^{-1/2}$ in Figure 5a. Constant behavior of $R_F$ irrespective of $f_{pump}$ indicates that $R_F$ is not dependent on the exciton density $n_0$, which is consistent with basic nature of $R_F$ represented by Eq. (5). In line with this, Figure 5b also shows that annihilation rate with FRET is almost independent of $f_{pump}$.

Third, we provide the evidence of FRET as an annihilation mechanism of A- and B-excitons by showing the spectral overlap between donor emission and acceptor absorption spectra. In case of homo-FRET, where the energy transfer occurs from the same entity, exciton emission and ESA are always overlapped as shown in Figure 6a. Efficient FRET can be expected for the annihilation step $E_1 + E_1 \rightarrow E_n + E_0$, with $E_n$ being the final state in the TA experiments. Here, $E_{10}$ is the energy of radiative recombination, photoluminescence of A-exciton, and $E_{12}$ is the energy of ESA from the first excited (exciton) state to second excited state. Depending on the pump energy, ESA can become $E_{1n}$, transition energy from the first excited (exciton) state to $n^{th}$ excited state. In general, $E_{10}$ is in between $E_{12}$ and $E_{1n}$ ($E_{12} < E_{10} < E_{1n}$), which was confirmed by previous work showing spectral overlap between the photoluminescence and the ESA of A-exciton [51]. Hence, this spectral overlap between $E_{10}$ and $E_{1n}$ is another critical evidence for FRET as an annihilation mechanism of A-exciton.

Finally, we corroborate the quenching of emission of the donor, or the reduction in lifetime of the donor due to its increased local density of optical states. We note that $\tau_1$, responsible for EEA, is strongly associated with the quenched lifetime of donor. In Figure 6b, significant decrease in $\tau_1$ with $f_{pump}$ for A-exciton (Table 1) is analogous to the decrease of donor exciton lifetime when FRET takes place. In order to understand more manifestly, we provide the schematic description of FRET between A-excitons (Figure 6c). Upon photoexcitation, multiple excitons are generated, and FRET occurs in such a way that energy is transferred between two excitons as one of two excitons recombines, reflected in a decrease in the exciton lifetime ($\tau_1 < \tau_2$). Energy transferred to another exciton leads to ESA, herein free carrier absorption in the conduction band. The hot electrons resulting from free carrier absorption eventually undergo intraband relaxation (usually the sub-ps time scale [33]) and interband relaxation (recombination of exciton, $\tau_2$). In here, decrease in $\tau_1$ manifestly shows the quenching effect of donor exciton.

## VI.  CONCLUSION

In conclusion, we have systematically studied the exciton annihilation mechanism in a $MoS_2$ monolayer based on the transient absorption spectroscopy. We observed the exciton-exciton annihilation of A-, B- and C-excitons and demonstrated that the multistep diffusion mediated by exciton-phonon scattering is responsible for C-exciton annihilation, while annihilation of A- and B-excitons is predominantly governed by direct FRET. We provided several critical evidences to underpin the FRET as an annihilation mechanism of A- and B-excitons: exclusion of the diffusion process, independence of Förster radius to temperature and exciton density, and donor quenching effect. Given that a comprehensive understanding of excitonic properties of monolayer $MoS_2$ is crucial for various applications in photonics and optoelectronics, this study will offer profound insight into the research on the excitonic properties of 2D TMDs.

**Acknowledgments**

This research was supported by Bill & Melinda Gates Foundation, National Science Foundation, and National Natural Science Foundation (NSFC, Grant Nos. 11804334).

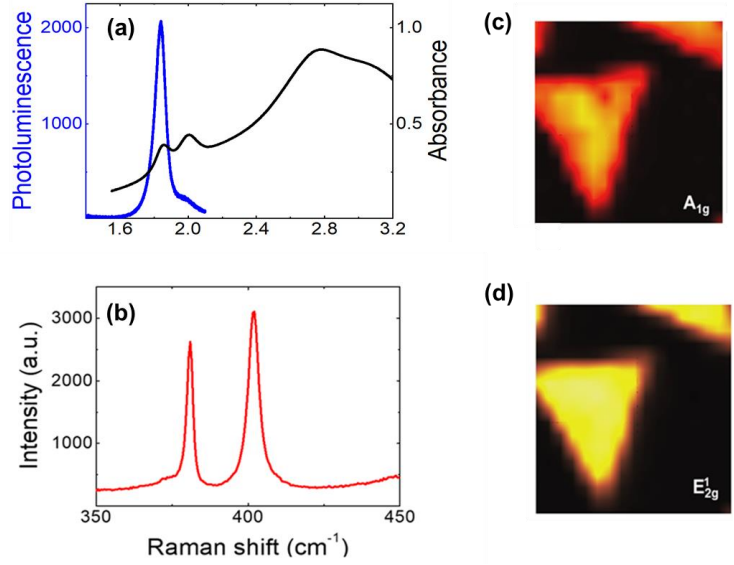

**Figure 1.** (a) Absorption and photoluminescence spectra of MoS$_2$ monolayers. (b) Raman spectrum of MoS$_2$ monolayer (c) and (d), Raman intensity map for $A_{1g}$ and $E^1_{2g}$ modes, respectively.

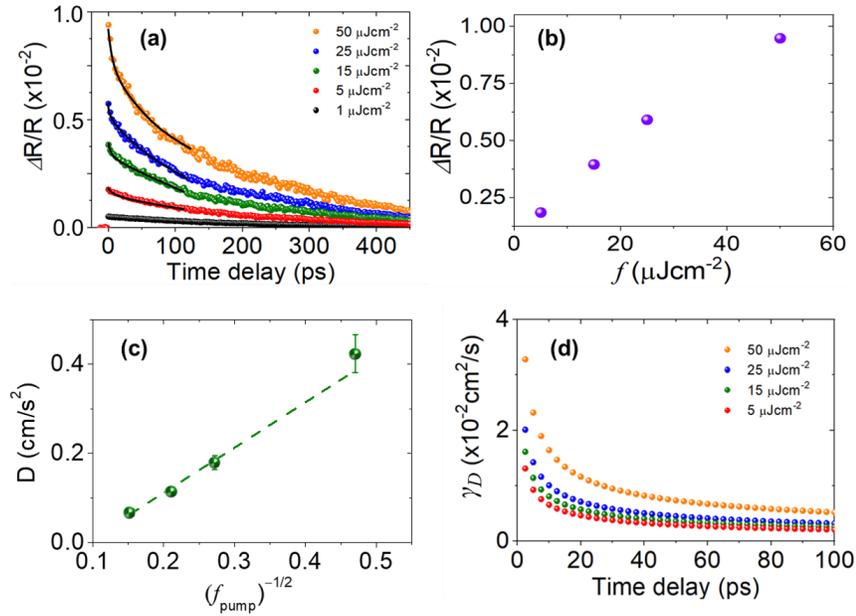

**Figure 2.** (a) Transient absorption kinetics for C-exciton for pump fluences of 1, 5, 15, 25, and 50 μJcm$^{-2}$ with fitting (back) curves up to ~ 100 ps based on exciton-exciton annihilation via diffusion. (Eq. (3)). (b) Initial amplitude of $\Delta R/R$ as a function of pump fluence. (c) Relationship between diffusion coefficient of C-exciton and inverse square root of pump fluence. (d) Plot of annihilation rate via diffusion as a function of time delay (Eq. (2a))

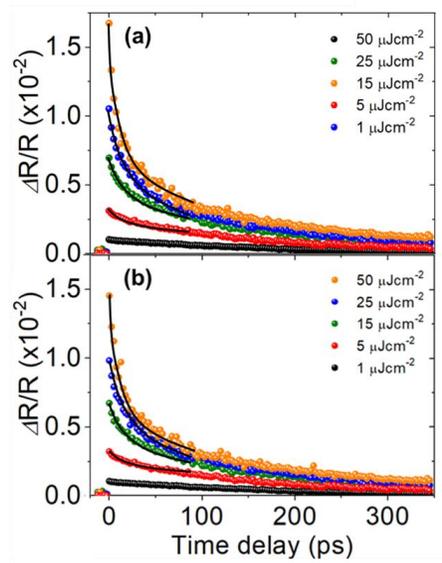

**Figure 3.** (a) Transient absorption kinetics for A-exciton and (b) B-exciton for pump fluences of 1, 5, 15, 25, and 50 µJcm$^{-2}$ with fitting (back) curves up to ~ 100 ps based on exciton-exciton annihilation via FRET. (Eq. (4))

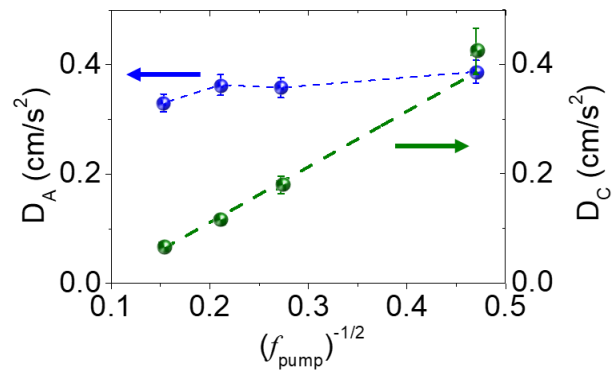

**Figure 4.** Plot of diffusion coefficient of A-exciton ($D_A$) and C-exciton ($D_C$) as a function of inverse square root of the pump fluence.

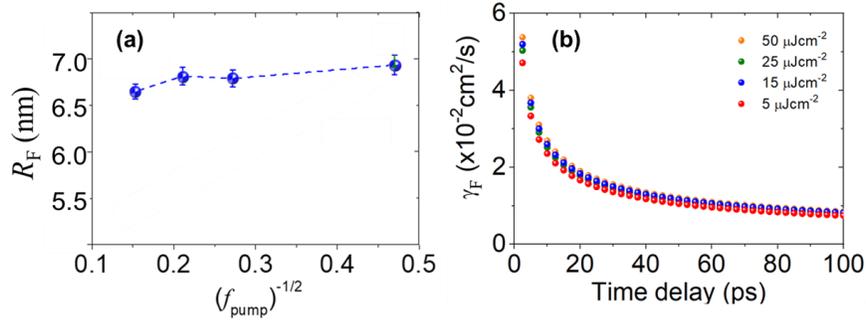

**Figure 5.** (a) Plot of Förster radius for A-exciton as a function of inverse square root of the pump fluence. (b) Plot of annihilation rate via FRET as a function of time delay (Eq. (2b))

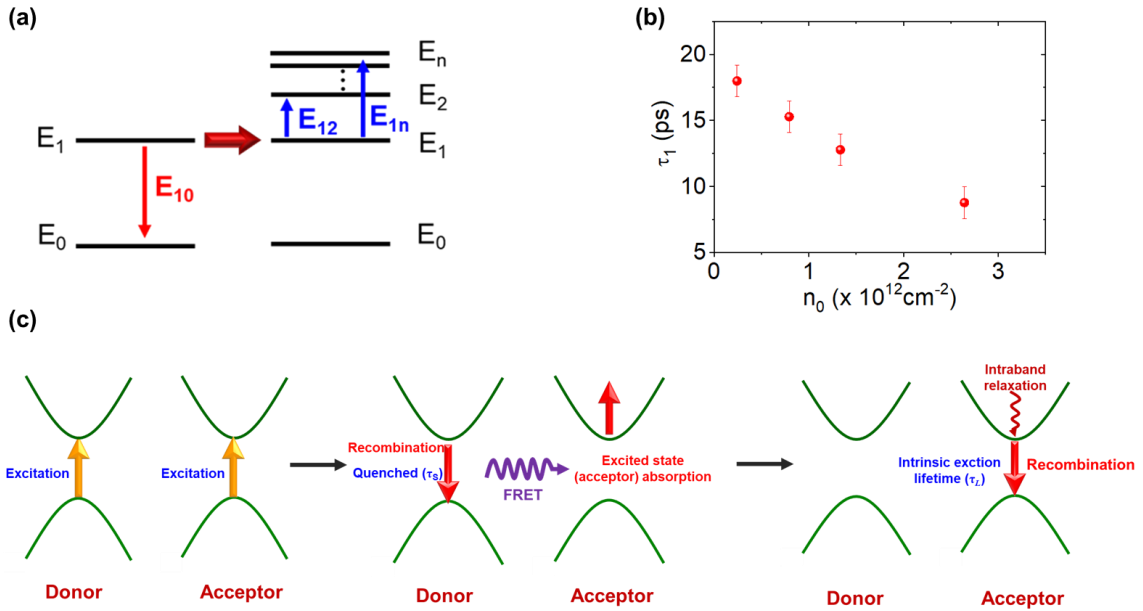

**Figure 6.** (a) FRET process between identical excitons. Exciton annihilation between two identical molecules can be described via the FRET with a spectral overlap between the exciton emission ($E_{10}$) and the absorption of the exciton state to higher electronic states ($E_{12}, ..., E_{1n}$). (b) Shorter decay time ($\tau_1$) as a function of initial exciton density. (c) Schematic description of FRET between A-excitons occurring in 2D MoS$_2$. Upon photoexcitation, multiple excitons can be generated simultaneously. Energy is transferred between two excitons as one of two excitons recombines non-

radiatively, resulting in quenching of donor exciton (decrease of $\tau_1$). Energy transferred to another exciton leads to free carrier absorption in the conduction band. The hot electrons undergo intraband or interband ($\tau_2$) relaxation.

| exciton / $f_{pump}$ (μJcm$^{-2}$) | A-exciton $\tau_1$ (ps) | A-exciton $\tau_2$ (ps) | B-exciton $\tau_1$ (ps) | B-exciton $\tau_2$ (ps) | C-exciton $\tau_1$ (ps) | C-exciton $\tau_2$ (ps) |
|---|---|---|---|---|---|---|
| 1 | – | 189 | – | 180 | – | 219 |
| 5 | 17.8 | 188 | 18.7 | 182 | 31.6 | 222 |
| 15 | 14.8 | 193 | 15.3 | 183 | 29.7 | 225 |
| 30 | 12.1 | 187 | 12.8 | 188 | 27.4 | 221 |
| 50 | 8.8 | 195 | 9.6 | 186 | 24.8 | 227 |

**Table 1.** Decay time constants of A-, B-, and C-excitons for several pump fluences.